# Topological balance of cell distributions in plane monolayers

Daria S. Roshal[1], Kirill K. Fedorenko[1], Marianne Martin[2], Stephen Baghdiguian[3,*], Sergei B. Rochal[1,**]

[1] Faculty of Physics, Southern Federal University, Zorge 5, Rostov-on-Don, 344090, Russian Federation
[2] VBIC, INSERM U1047, University of Montpellier, Montpellier, 34095 France
[3] Institut des Sciences de l'Evolution-Montpellier, Université de Montpellier, CNRS, Ecole Pratique des Hautes Etudes, Institut de Recherche pour le Développement, Montpellier, 34095 France

* Corresponding author. E-mail: stephen.baghdiguian@umontpellier.fr
** Corresponding author. E-mail: rochal_s@yahoo.fr

## Abstract

Most of normal proliferative epithelia of plants and metazoans are topologically invariant and characterized by similar cell distributions according to the number of cell neighbors (DCNs). Here we study peculiarities of these distributions and explain why the DCN obtained from the location of intercellular boundaries and that based on the Voronoi tessellation with nodes located on cell nuclei may differ from each other. As we demonstrate, special microdomains where four or more intercellular boundaries converge are topologically charged. Using this fact, we deduce a new equation describing the topological balance of the DCNs. The developed theory is applied for a series of microphotographs of non-tumoral epithelial cells of the human cervix (HCerEpiC) to improve the image processing near the edges of microphotographs and reveal the topological invariance of the examined monolayers. Special contact microdomains may be present in epithelia of various natures, however, considering the well-known vertex model of epithelium, we show that such contacts are absent in the usual solid-like state of the model and appear only in the liquid-like cancer state. Also, we discuss a possible biological role of special contacts in context of proliferative epithelium dynamics and tissue morphogenesis.

## 1. Introduction

Epithelial tissues play an important role in the functioning of metazoans, protecting their internal organs from damage and transporting nutrients. Despite the fact that the epithelia with polygonal intercellular boundaries are known for several centuries, it is still under discussion how cellular geometry affects mechanical properties of epithelia, processes of cell growth, division and death [1-5]. Moreover, transformation of flat epithelial monolayers into structures with complex geometry during embryonic development can also be connected with some geometrical changes in the tissues [6-11].

In spite of the fact that the polygonal structure of the onion epithelium was discovered by R. Hooke back in 1665, the first statistical analysis of the cell distribution according to the number of their nearest neighbors (DCN) was carried out in 1926 [12]. Much later, in 2006, it was shown that normal proliferative (slowly dividing) epithelia of many plant and animal species are characterized by similar DCNs [2,13-18]. This topological invariance is closely related to the physiological invariance of normal epithelium: in all metazoans the epithelium acts as a selective barrier that controls the fluxes of nutrients, regulates ionic and water

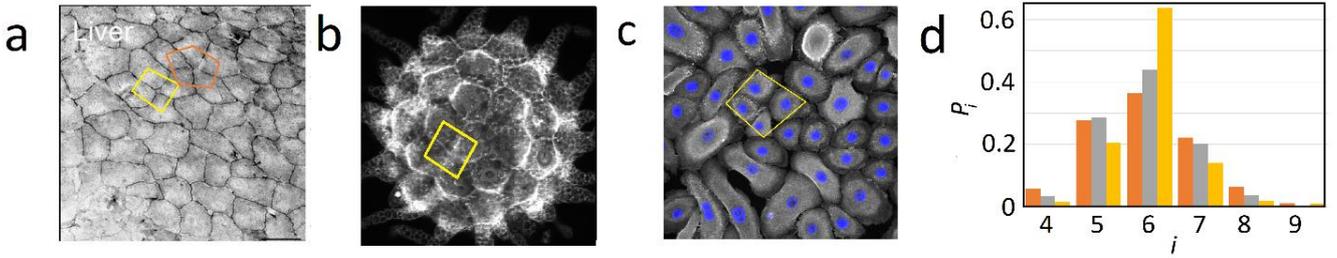

**Figure 1.** Specific features of the cell convergence in epithelial monolayers. (a) Liver cells. Image adapted from work [30]. Convergences of boundaries of four and five cells are marked in yellow and orange. (b) Follicular epithelial cells on the surface of an oocyte of *Ciona intestinalis*. (c) Human cervical cell (HCerEpiC) monolayer. Cell nuclei are stained with blue dye. Microphotographs (b-c) were previously obtained by us in [26,28]. Yellow quadrilaterals bound four cells that converge at one point. (d) Distribution of cells by number of neighbors (DCN) for epithelia of different types. DCNs for Human cervical cancer (HeLa), and normal (HCerEpiC) cells, and follicular epithelium of ascidia oocytes are shown with orange, grey and yellow [26,28].

balance, and protects the host from antigens and microbes [19-20].

DCN can be used for assessing biological properties of epithelial tissues. For example, it is known that specific topological defects in epithelium can be precursors of apoptosis [21]. Topological analysis of epithelial monolayer can be used to diagnose cancer, as well as to study morphogenesis processes [22-25]. In particular, in our work [26] it was shown that monolayers of cancerous (HeLa) and healthy (HCerEpiC) cells of the human cervix may be distinguished by the proportion of cells that have 6 neighbors. In cancer monolayers, the cell distribution is close to random one, the cells are more elongated, and the proportion of hexavalent cells is less than in normal epithelium. In contrast, in many non-proliferative epithelia in which cell division has already ceased, such as in Drosophila wings or ascidian oocytes (see DCNs for epithelia of different nature shown in Fig. 1 (d)), the proportion of hexavalent cells is greater than in normal epithelia [27-29]. At the same time, according to the results [30] obtained using computer vision, in the non-proliferative visceral organ surface epithelia the proportion of hexavalent cells turns out to be significantly smaller.

However, computer algorithms used to analyze microphotographs of epithelium can make systematic errors associated with the complexity of processing both the edges of microphotographs and blurred images of nuclei or cell boundaries. Note that a convergence of three cell boundaries at one point is typical of epithelia, but sometimes four, or less often five cell boundaries converge at one microdomain (see Fig. 1), which can also lead to difficulties in determining the number of nearest neighbors (valency) of cells.

If cell nuclei are visible in micrographs, then to determine the cell valences it seems quite natural to use the centers of these nuclei as nodes of the Voronoi tessellation and determine the number of cell neighbors by the number of sides of the resulting polygons. Recall that a Voronoi polygon is defined as a boundary of a set of points that are closer to the considered node than to any other. It is extremely unlikely to find a symmetrical arrangement of nuclei in a cell monolayer, therefore the features of the location of cell boundaries shown in Fig. 1 disappear in the Voronoi tessellation, and exactly three polygons converge at each vertex of this tiling. In this case, the DCN for a sufficiently large monolayer turns out to be balanced [29], namely

$$\sum_i p_i q_i = 0, \qquad (1)$$

where $p_i = N_i/\Sigma_j N_j$ is the proportion of cells with $i$ neighbors; $N_j$ is the number of cells with $j$ neighbors; $q_i = 6 - i$ is the topological charge of a cell with $i$ neighbors. Note that, as our analysis shows, for experimental DCNs constructed using different methods in [12-14], the right-hand side of (1) differs from zero within 0.05, but in [30] this imbalance turns out to be 3 times greater.

Is this a consequence of incorrect operation of algorithms based on machine learning, or is Eq. (1) for epithelial monolayers discussed in [29] inapplicable? What to do if it is impossible to identify the geometric center of a cell, and yet there is a need to determine the valence of cells by the number of its sides? Does the formula (1) remain applicable when the fraction of microdomains at which more than 3 cell boundaries converge is significant? Thus, the aim of this article is to answer these questions, generalize expression (1) and discuss the method of finding DCN using a series of microphotographs of human cervical cell (HCerEpiC) monolayers.

## 2. Balance of the distribution of polygon types in a nondegenerate tessellation of the plane

If in a Voronoi tessellation exactly three polygons converge at each vertex, then such a tiling is *uniquely* equivalent to a triangulation whose edges are perpendicular to the sides of the polygons. It is easy to show that for any triangulation of an infinite plane, the average number of edges converging to one node is 6. To do this, we can use a generalization of Euler's theorem that relates the number of edges ($E$), polygons ($P$) and vertexes ($V$) of the tessellation of the plane into polygons: $V - E + P = 0$. In this case, the polygons are triangles bounded by three edges, so $3P = 2E$ and $V = E/3$. Since each edge has two ends, then, on average, each vertex coincides with 6 ends of edges. Thus, both for a nondegenerate Voronoi tessellation and for its dual triangulation, Eq. (1) is true.

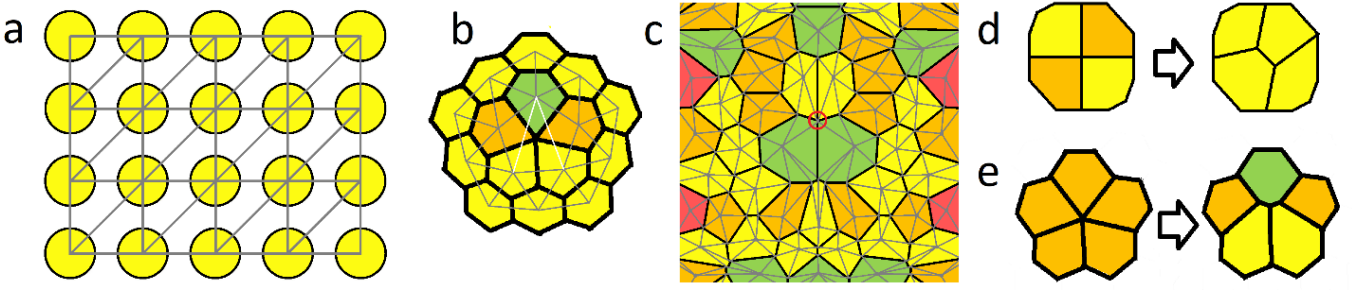

**Figure 2.** Topological features of a degenerate polygonal tessellation. (a) An example of an ambiguous triangulation of a square packing, in which each vertex of the triangulation has 6 neighbors. (b) An ambiguous part of triangulation (two lines in white) corresponds to the region containing a group of five cells. (c) A degenerate polygonal tessellation and an auxiliary triangulation superimposed on it with vertices coinciding with both the vertices and centers of the polygons. The degenerate contact is circled in red. (d-e) Ways to remove the degeneracy of the contact of 4 and 5 polygons. The color of circles and polygons in (a-b) corresponds to the number of neighbors according to shown triangulations. The color of polygons in panels (c-e) denotes the number of polygon sides. Red, orange, yellow and green encode the numbers 4,5,6 and 7 respectively.

Moreover, Eq. (1) is valid for any polygonal tessellation of the plane if for this tessellation there is a dual triangulation for which its nodes and edges uniquely correspond to the centers of polygons and their edges, respectively. Note, however, that for a finite region of even a nondegenerate tessellation (e. g. micrograph of a sample), Eq. (1) becomes approximate. In what follows we modify Eq. (1) for the case of degenerate tessellation and consider a way to decrease an error associated with the microphotograph boundaries.

## 3. Balance relation for degenerate tessellation

A simple example of a degenerate polygonal tessellation is a square packing, in which 4 edges converge at each vertex, and which therefore violates relation (1). However, if you connect the vertices of the packing by any triangulation, then the average number of neighbors by triangulation will be equal to 6 and relation (1) will remain the same, see the example of triangulation in Fig. 2(a). In the same ambiguous way, a region containing a group of four or five cells (see Fig. 2(b,c)) with a common point of contact can be triangulated. Further, if necessary, one can reconstruct the polygonal tessellation so that it becomes nondegenerate and thus restore relation (1) (see Fig. 2(d,e)). Note that if this reconstruction occurs in a finite region than a total topological charge of polygons and convergence points located in the region is conserved. For example, for the cases shown in Fig. 2(d,e) these charges are 0 and +1, respectively.

Let us obtain the balance relation for a polygonal tessellation which contains degenerate contacts. Let's introduce a center for each of the polygons, so that this point can be connected by line segments with all vertices of the polygon. In the simplest case, the center of the polygon can be chosen in such a way that the additional segments completely belong to the polygon. Then it is obvious that there is an auxiliary triangulation, the nodes of which are both the centers of the polygons and their vertices. For the case of strongly concave polygons, for which it is impossible to introduce the center in this way, the existence of auxiliary triangulation is not obvious. However, when analyzing various cell monolayers, we have never encountered such a cellular configuration; here we will not consider the validity of our analysis for the general case.

Next, we associate topological charges with all nodes of the auxiliary triangulation in the usual way. Then the charge $q$ of the center of a polygon with $n$ sides is equal to $q = 6 - n$, and the charge of its vertices is $Q = 6 - 2m$, where $m \geq 3$ is the total number of polygon edges converging to this vertex. The number $m$ doubles since the edges of the introduced auxiliary triangulation also contribute to the vertex valences. The total topological charge of all nodes of the auxiliary triangulation for an infinite tessellation must be equal to zero. Therefore

$$N \sum_i p_i(6 - i) = -\sum_j Q_j, \qquad (2)$$

where $N = \Sigma_j N_j$ is the total number of polygons, the left side of (2) corresponds to the topological charge of their centers, and this summation is carried out according to the types of polygons in the DCN. On the right side with a minus sign there is a topological charge created by the degenerate vertices of the polygonal tessellation. Thus, we obtain the generalized balance equation:

$$\sum_i p_i(6 - i) = \sum_{l>3} R_l(2l - 6), \qquad (3)$$

where $R_l$ is a ratio between the number of vertices at which more than three edges of polygons converge and the total number of polygons. For a nondegenerate polygonal tessellation, the right side of (3) vanishes and relation (3) turns into (1).

As a simple demonstration of the relations (2-3) validity, we note that according to these relations, an infinite planar square packing is completely balanced: each square carries a topological charge $q = +2$, and each vertex $Q = -2$. Since Eqs. (2-3) are exact for an infinite planar tessellation, they can be used to check the results of obtaining DCNs from experimental data.

Below, using an example of a small series of micrographs of Human cervical cell (HCerEpiC) monolayers [26], we apply the above-established topological relations for the data analysis.

## 4. Analysis of a series of micrographs of HCerEpiC monolayers

This series consists of 21 micrographs containing from 30 to 54 cells. The method for obtaining the cell monolayers and their microphotographs is described in Method section. The fact that microphotographs contain a small number of cells reduces the accuracy of determining the DCN but allows us to develop a method to consider correctly the image boundaries. In the photographs both cell nuclei and cell boundaries are clearly visible, however, it is very difficult to distinguish the fine structure of cell contacts, so special points at which three or more cell boundaries converge were located visually. Further computer processing in order to obtain the cell valences based on the location of their boundaries is actually amounted to the transition from real boundaries to polygons approximately surrounding the cells. Obviously, if the polygon vertices coincide with the special points on the cell boundary, then the number of the polygon sides is equal to the cell valency. We note that some sides of polygons can strongly intersect the cells with a large curvature of the border, but this fact does not matter for calculating the cell valences.

In Fig. 3(a) only those polygons that correspond to cells with fully visible boundaries are fully constructed. A simple way to obtain the DCN is to take into account all such polygons in all images. However, checking the distribution constructed in this way leads to a discrepancy between the left and right sides of (3) by approximately 0.19. First, this is due to the fact that the images contain not large numbers of cells. Second, small cells, which according to [12] have a small number of neighbors, are more likely to be *completely* fitted in the image, even if they locate near the boundaries of the micrograph. To level out the last fact, additional processing of the edges of the microphotographs is done. On each image a rectangle is drawn in such a way that it did not include a single point of polygons with unknown numbers of sides, and so that the area of such a rectangle was as large as possible. For example, the lateral and bottom boundaries of the rectangle highlighted in red in Fig. 3(a) are drawn through the vertices (shown in orange and brown) of the outer polygons, for which the number of sides cannot be accurately determined. The upper boundary is drawn through the special point corresponding to the vertex of the polygonal boundary surrounding the partially visible large cell.

The yellow histogram in Fig. 3(b) shows the DCN based only on those polygons that at least partially located inside the rectangles highlighted in the microphotographs of the analyzed series. The distribution imbalance according to Eq. (1) is equal to $\sum_i p_i q_i \approx 0.81$. However, for the same distribution, the discrepancy between the left and right sides of expression (3) is only $\sim -0.01$, which indicates a fairly good quality of experimental data processing.

Next, we decided to consider how DCN will change if all degenerations of cell contacts are forcibly removed. To do this, it is necessary to distribute the topological charges, that carry degenerate cell contacts, over neighboring cells. If we formally assume that the topological charge is distributed uniformly when degeneracy is removed, then cells neighboring with $l$-fold contact point decrease their topological charges by value of $Q_{add} = (2l - 6)/l$, which formally corresponds to an increase in the number of neighbors by the same quantity. The resulting effective valences of cells are shown in Fig. 3(a) with green numbers.

To take into account in the DCN a cell that has a non-integer number of nearest neighbors of the form $i + f$, where $i$ is the integer part of the number, $f = n/d$ is the fractional part (presented as the ratio of two integers $n$ and $d$) we consider such a cell as $n/d$ cells with $i$ neighbors and $1 - n/d$ cells with $i + 1$ neighbors. The DCN calculated in this way is presented in Fig. 3(b) with a green histogram. This distribution is well balanced: $\sum_i p_i q_i \approx -0.01$.

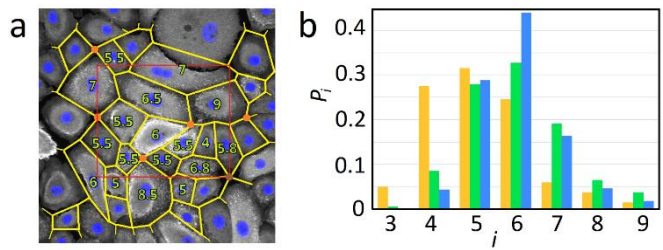

**Figure 3.** Analysis of microphotographs of HCerEpiC monolayers. (a) Vertices of the yellow polygons correspond to points of concurrent contact of 3 or more cells. In the latter case, the contact points carry a topological charge; contacts of four and five cells are marked with orange and brown circles, respectively. For cells located within the selected zone, shown by the red rectangle, green numbers indicate their effective non-integer topological charges (see the main text). (b) Cell distributions according to the number of their nearest neighbors (DCNs) obtained by three different methods described in the text.

For this series of micrographs, we also calculated the DCN based on the Voronoi tessellation with nodes located at the centers of cell nuclei (see the histogram shown in blue in Fig. 3(b)). It is almost impossible to encounter a locally degenerate arrangement of nuclei (more than three nuclei equidistant from their common center) in cell monolayers, therefore the Voronoi tessellation always turns out to be nondegenerate. To determine correctly the DCN in the case of using Voronoi tessellation, one needs to be sure that there are no nuclei outside the image boundary that can change the valence of those nuclei that are taken into account when calculating the DCN. Therefore, the center of a reliably constructed Voronoi polygon should be located at least twice as far from the image border as any of the vertices of this polygon. However, this way, for the same reasons as constructing the DCN using all visible polygons, leads to the appearance of excessive total positive topological charges localized at the image border. As a result, the imbalance of the DCN $\sum_i p_i q_i$ takes the value of $\sim 0.1$.

To improve the quality of experimental data processing and reduce the imbalance, when constructing the DCN, we take into account only those Voronoi polygons whose centers are located inside the rectangles selected in the following way. The rectangles should not contain nodes with undetermined valence (cut off in the previous step), while their area should be as large as possible. Thanks to this processing of image edges, the imbalance of the DCN is reduced by half. The corresponding histogram in Fig. 3(b) is shown in blue.

## 5. Discussion and conclusion

HCerEpiC cell monolayers are a typical representative of normal proliferative epithelium, and if we compare the different methods of processing microphotographs presented above, then processing using the construction of the Voronoi tessellation most correctly reflects the following fact: in different epithelia of this type (like Drosophila wing disc, Xenopus tail epidermis or Hydra epidermis) $P_6$ changes from 0.42 to 0.48 [12]. For the case of HCerEpiC monolayers, when the calculations are based on intercellular boundaries (see the orange histogram in Fig. 3(b)), the high $P_4$ value and decrease in the $P_6$ one are explained mainly by the topological charges carried by degenerate contact points. In this context we note that the DCNs obtained in refs. [12-14] are well balanced, i.e the inaccuracy of Eq. (1) is less than 0.05. It means that degenerate contacts in these normal monolayers are very rare, and this fact can be confirmed by direct visual analysis of published images of corresponding epithelia.

The difference between green and blue histograms is mainly caused by the chosen way to transfer the topological charges when the contact degenerations are lifted. Obviously, there must exist small shifts in the intercellular boundaries that make both histograms very similar. It should also be noted that when one uses the Voronoi tessellation, a slight averaging of the cell areas occurs; i.e, on average, the areas of large cells decrease, and small ones increase, which also leads to an increase in the value of $P_6$.

Next, it is of interest to briefly discuss the biophysical mechanisms that can lead to the appearance of degenerate contact points between cells. The first of them, in our opinion, is purely biological, and can work in proliferative monolayers, in which planar polarization is closely linked to the presence of intercellular contacts, or channels between adjacent cells [31]. Then if a pair of such cells divides at approximately the same time, then there is a high probability of the appearance of a degenerate contact point between four cells at once.

In addition, we recall that the cortex is a polymerized actin-rich zone located just beneath the membrane of animal cells. In association with myosin, it is involved in cell dynamics and shape. Myosin generates mechanical stresses in the actin network, which therefore behaves as an active material at the origin of spatially localized actin-enriched structures [32]. The degenerate contacts observed in epithelial structures could therefore locally increase actin concentration and be an intercellular mechano-chemical signaling platform required for the dynamics of proliferative epithelia and tissue morphogenesis.

The second possible mechanism is biophysical and is associated with minimizing the area of intercellular boundaries. Such minimization can lead to an attraction between special points on the boundaries and ultimately cause the appearance of degenerate contacts. We do not construct our own model of this phenomenon but limit ourselves by consideration of the possible appearance of degenerate contacts in one of the most popular models of epithelia: vertex model [33,34].

In this model, the centers of cells are considered as nodes of some Voronoi tessellation. The energy $E$ of an epithelium containing $N$ cells reads:

$$E = \sum_{i=1}^{N} \beta (A_i - A_0)^2 + \zeta (P_i - P_0)^2, \qquad (4)$$

where the subscript $i$ labels each cell; $A_i$ and $P_i$ are the cell areas and perimeters, respectively. The origin of the elastic modulus $\beta$ is associated with a combination of the cell volume incompressibility and resistance to height differences between nearest cells [34,35]. The second term including the cell perimeters $P_i$ results from active contractility of the actomyosin subcellular cortex and effective cell membrane tension due to cell-cell adhesion and cortical tension, which are linear in perimeter. To obtain a planar model monolayer, a certain set of nodes is first specified, and then using periodic boundary conditions, energy (4) is minimized relative to the coordinates of these nodes.

The model has two states with different mechanical properties [34,35]. The so-called jamming transition between the states is controlled by the target shape index $s_i = P_0/\sqrt{A_0}$ [34,35]. In the glass-like state $s_i < s_0$, while in the fluid-like one $s_i > s_0$, where $s_0 \approx 3.81$ [34,35]. In the liquid state the cells take more elongated shape, and the fraction of 6-valent cells decreases [34,35]. Please, see more detailed description of the vertex model in refs. [33-35].

Having analyzed more than 100 model monolayers obtained at different values of the parameters $s_i$ and ratio $\beta/\zeta$, we found that in the glass-like state (see Fig. 4(a)), the degeneration of cell contacts is observed quite rarely and is approximate. This makes the model monolayers obtained similar to normal proliferative epithelia considered in refs. [12-14]. In the fluid-like state (see Fig. 4(b)), this degeneration is observed much more often, and within the accuracy of numerical calculations, the degeneration is not lifted. For example, Fig. 4(b) shows a fragment containing 60 Voronoi polygons, in which there are 8 degenerate contacts between four polygons and 5 degenerate contacts between five or more polygons. In a fluid-like state, the polygon perimeter is redundant, and the appearance of degenerate contacts makes it possible to minimize the energy (4) at this constraint.

Within the model, a normal proliferative monolayer is usually interpreted as a glass-like state; in this state, the appearance of degenerate contacts in the model is random and not regular. While this is true for proliferative epithelia examined in ref. [12-14], the vertex model, however, cannot describe the occurrence of degenerate contacts between cells either in HCerEpiC monolayers or in the monolayers studied in [30]. In contrast, in the fluid-like state that corresponds to a cancerous (hyperproliferative) epithelium [26], the model readily leads to the appearance of degenerate contacts between cells and one can find them in published images [36] of human colon.

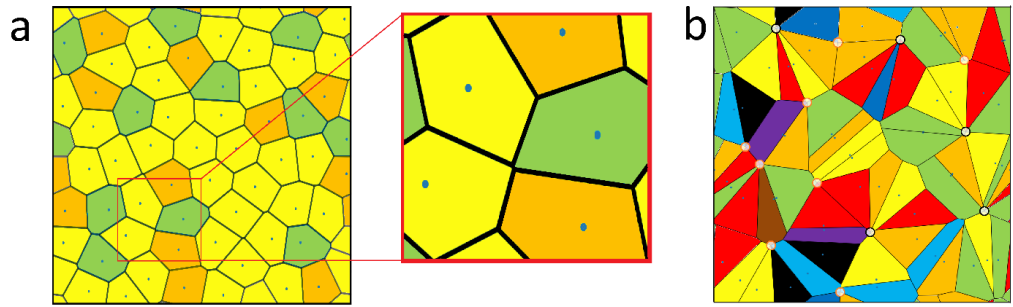

**Figure 4.** Degeneration of cell contacts within the vertex model of epithelium. (a) Glass-like state of the model. In the center of the insert there is a contact where boundaries of four cells approximately converge. If you zoom in on the image, a short border between cells becomes noticeable. (b) Fluid-like state of the vertex model. Orange and black open circles indicate contacts of four and more than four polygons, respectively. The color of the polygons (black, red, orange, yellow, green, cyan, indigo, violet, brown) corresponds to the number of nearest neighbors, which varies from 3 to 11.

In conclusion, in this Article we study the topological characteristics of cell distributions according to the number of cell neighbors and obtain a balance equation for the considered distributions. If there are special points in the epithelium, where 4 or more cells contact, then such points carry a topological charge and contribute to the balance equation. Biophysical mechanism generating the special points is also discussed. Our article may be useful for researchers studying the relations between the geometric characteristics of the epithelium and its biophysical properties.

## 6. Method

*Cell line growth and fluorescence confocal microscopy*

Normal primary cervical epithelial cells (HCerEpiC) isolated from human uterus were purchased from ScienCell Research Laboratories (Clinisciences S.A.S., Nanterre, France). Cells were grown, processed for fluorescence microscopy and examined under a Leica TCS SPE confocal microscope as previously described [26]

## Acknowledgements

D.R. and K.F. acknowledge financial support from the Russian Science Foundation, grant No. 22-72-00128.